\documentclass[12pt]{article}
\usepackage{amsmath}
\usepackage{amssymb}
\usepackage{bm}
\usepackage[dvips]{graphicx}
\usepackage{color}

\begin{document}

\begin{titlepage}

 \setcounter{page}{0}

 \begin{flushright}
  KEK-TH-1178 \\
  OIQP-07-11 \\
  YITP-07-56
 \end{flushright}

 \vskip 5mm
 \begin{center}
  {\Large\bf  Hawking Radiation via Higher-spin Gauge Anomalies} 

  \vskip 15mm

 {\large
 Satoshi Iso$^{1,}$\footnote{\tt satoshi.iso@kek.jp},
  Takeshi Morita$^{2,}$\footnote{{\tt mtakeshi@yukawa.kyoto-u.ac.jp,
  takeshi@theory.tifr.res.in}, \\
  Present address: Department of Theoretical Physics, Tata Institute of
  Fundamental Research, Homi Bhabha Road, Mumbai 400005, India. 
  }  
 and Hiroshi Umetsu$^3,$\footnote{\tt hiroshi\_umetsu@pref.okayama.jp}
  } 

  \vspace{5mm}
  $^1${\it Institute of Particle and Nuclear Studies \\
  High Energy Accelerator Research Organization (KEK) \\
  Oho 1-1, Tsukuba, Ibaraki 305-0801, Japan 
  }
 
  \vspace{5mm}

  $^2${\it Yukawa Institute for Theoretical Physics, Kyoto University
  \\
  Kyoto 606-8502, Japan}
  
  \vspace{5mm}
  $^3${\it Okayama Institute for Quantum Physics \\
  Kyoyama 1-9-1, Okayama 700-0015, Japan} 
 \end{center}

\vskip 30mm 

 \centerline{{\bf{Abstract}}} 

 \vskip 3mm 
 We give a higher-spin generalization of the anomaly 
 method for the Hawking radiation from black holes.
In the paper \cite{IMU4} higher-spin  generalizations of the 
 gauge (and gravitational) anomalies 
 in d=2 were obtained.
 By applying these anomalies to black hole physics, 
 we derive the higher moments of  the Hawking fluxes. 
We also give a higher-spin generalization of 
the trace anomaly method by Christensen and Fulling \cite{CF}.
\end{titlepage}

\newpage

\section{Introduction}
\setlength{\baselineskip}{7mm}

\setcounter{footnote}{0}
\setcounter{equation}{0}
Hawking radiation is a universal quantum effect which arises in the
background spacetime with event horizons \cite{Hawking1,Hawking2}. 
The radiation is essentially originated from the causal structure of  the
event horizon and the quantum behavior of fields around it.
Such universal behavior is  emphasized  in the recent derivation of 
Hawking radiation based on gravitational anomaly \cite{Robinson}. 
One of the key observations is that near the horizon,
due to the infinite blue shift, 
ordinary  matter fields in the black hole backgrounds behave as if they
are an infinite set of  massless two-dimensional fields, 
and the causal structure of the horizon  
makes the two-dimensional fields effectively chiral at the horizon. i.e.
the ingoing modes (which can be interpreted as left moving
modes) can be classically decoupled from the outside region 
of the horizon. Chiral theories suffer from gravitational 
 anomalies.
Since the underlying theories must be covariant,
they should be cancelled by the quantum effect of 
classically irrelevant ingoing modes at the horizon.
The requirement of the  anomaly cancellation 
 leads to the existence of the
Hawking energy-flux from the black hole
in the Unruh vacuum \cite{Unruh}. 
This method was, after correcting some errors in    
the original procedure\footnote{It is explained in the appendix \cite{IMU1}
why the original  procedure in \cite{Robinson} 
does not work for charged or rotating  black holes},
generalized to 
charged black holes \cite{IUW1},
 and further applied to rotating black holes \cite{IUW2, Soda}
 and others \cite{IMU1,others}.

The  anomaly method, so far, could 
obtain only partial information of 
Hawking radiation because the gravitational and gauge anomalies
contain only the information of fluxes of energy and charge. 
These fluxes correspond to the zero-th and first moments
of the thermal distribution of radiation, and 
in order to obtain the full information, 
we need to calculate  all the other higher moments. 
In \cite{IMU2}, by using conformal field theory technique,
we calculated fluxes of higher-spin
(HS) currents 
in a Schwarzschild black hole.
They correspond to the higher-moments of Hawking radiation
and   the complete thermal distribution is reproduced. 
In this calculation we have used 
transformation properties of HS currents under
conformal transformations, which relate quantities near the horizon with
those at infinity. 
This method was further generalized
to charged black holes by considering gauge transformations in addition to
the conformal transformations \cite{IMU3}.
It is, however, not a direct generalization of the anomaly
method and the relation to the previous approaches
such as the anomaly method \cite{Robinson, IUW1}
or the method by Christen and Fulling \cite{CF}
 was not clear. 

In this paper, by using the  generalizations 
of gauge (and gravitational) anomalies in d=2 \cite{IMU4}, 
we show that
we can reproduce the  higher-spin fluxes of Hawking 
radiation for spin 3 and spin 4 currents constructed from 
fermions. This is the direct generalization of the 
anomaly method \cite{Robinson, IUW1}.
We also show that by solving the conservation equations and
trace anomalies for spin 3 and 4 currents, we can again 
reproduce the same fluxes. 
It is a HS generalization of the trace anomaly method by
Christensen-Fulling \cite{CF}.
These calculations
 can be straightforwardly extended to higher spins than 4,
 but they become very complicated.
 Instead of it, we show how we can obtain the asymptotic
 and near-horizon
 behavior of HS currents by using the results in \cite{IMU4}.

The organization of the paper is as follows.
In section 2, we first summarize the results
of higher-spin gauge and trace anomalies obtained in 
\cite{IMU4}. In section 3, we use the results
of HS gauge anomalies (up to spin 4)
to reproduce the higher spin fluxes of Hawking radiation.
Then in section 4, we  give a direct generalization
of the method by Christensen and Fulling.
Finally, by using the relations between 
holomorphic currents and covariant currents in \cite{IMU4}, 
we reproduce the asymptotic and near horizon behavior of 
the HS currents, and obtain
the full thermal spectrum of Hawking radiation.
These derivations  clarify some points which were obscure in the
previous papers \cite{IMU2,IMU3}. 

\section{Higher-spin gauge and trace anomalies}
\setcounter{equation}{0} 
\label{anomalies} 

In this section, we briefly  summarize 
anomalies for higher-spin currents
obtained in \cite{IMU4}.  We consider a two-dimensional
system with $c_L$ left-handed and $c_R$ 
right-handed fermions in the electric and gravitational 
backgrounds.  
If $c_L \neq c_R$,  the gauge 
and the general coordinate invariance are broken
due to the gauge and gravitational anomalies.
In the paper \cite{IMU4}, we showed that the higher-spin
gauge currents constructed from the fermions
also receive quantum anomalies in such chiral theories. 
Since we are here interested in the application to
the Hawking radiation from charged black holes,
it is sufficient to know higher-spin generalizations
of the gauge (or gravitational) anomalies 
in the electric and the gravitational backgrounds, i.e.
no HS gauge field backgrounds are necessary.

For the $U(1)$ current, it is well known that the gauge
and  chiral anomalies are given by
\begin{align}
 \nabla_\mu J^{\mu}
& = -\frac{(c_R-c_L)}{2}\frac{ \hbar}{2\pi} \epsilon^{\mu\nu}  F_{\mu\nu} ,
\label{conservation spin 1}\\
 \nabla_\mu J^{5\mu}
& = \frac{(c_L+c_R)}{2}\frac{ \hbar}{2\pi} \epsilon^{\mu\nu}  F_{\mu\nu}.
\end{align}
(\ref{conservation spin 1}) is  the gauge anomaly 
of the covariant current in d=2. 

In the case of the energy-momentum (EM) tensor, 
we have the gravitational
and trace anomalies,
\begin{align} 
\nabla^\mu T_{\mu\nu}=& F_{\mu\nu} J^\mu- \frac{\hbar}{48\pi} \frac{c_R-c_L}{2} \epsilon_{\mu \nu} \nabla^\mu  R,  
\label{conservation spin 2}
\\
{T^\mu}_{\mu}=&\frac{\hbar}{24\pi} \frac{c_L+c_R}{2}R.  
\end{align} 
This is  the covariant form of the gravitational anomaly
for the covariant EM tensor.
The first term of the r.h.s. in (\ref{conservation spin 2})
 is  classical 
violation of the matter energy momentum tensor
in the electric background by the coupling
between the charge current $J$ and the electric field $E$.
 This term arises because
the electric background transforms under the general
coordinate transformations.

These anomalies for the $U(1)$ current and the EM tensor
are usually calculated by the diagrammatic method
 or in the path integral formulation \cite{Fujikawa, Bartleman},
 but in d=2 they can be also calculated  easily 
 by using conformal field theory technique.
In \cite{IMU4}, we generalized this CFT method to calculate
anomalies for spin 3 and 4 currents 
constructed from  fermions. 

In the case of spin 3 current, the corresponding
 gauge and trace anomalies 
are given by 
\begin{align}
 \nabla_\mu {J^{(3)\mu}}_{\nu\rho}
 =&-  F_{\nu\mu} {J^{(2)\mu}}_\rho-  F_{\rho\mu} {J^{(2)\mu}}_{\nu}
 -\frac{1}{16}\nabla_\nu\left(R  J^{(1)}_\rho \right)  
 -\frac{1}{16}\nabla_\rho\left(R  J^{(1)}_\nu \right)
 + \frac{1}{16}g_{\nu\rho}\nabla_\mu \left(R J^{(1)\mu} \right) \nonumber \\
 &+ \frac{\hbar}{48\pi}\frac{c_R-c_L}{2}\left(  \epsilon_{\nu \sigma} \nabla^\sigma \nabla_\mu {F^\mu}_{\rho} + \epsilon_{\rho \sigma} \nabla^\sigma \nabla_\mu {F^\mu}_{\nu}- g_{\nu\rho}   \epsilon_{\alpha \sigma} \nabla^\sigma \nabla_\mu F^{\mu \alpha}  \right),
 \label{conservation spin 3} \\
{J^{(3)\mu}}_{\mu \nu} =& \frac{\hbar}{12\pi}\frac{c_L+c_R}{2}  \nabla_\mu {F^\mu}_{\nu}.
\label{trace spin 3}
\end{align} 
See \cite{IMU4} for the details of the definitions of the spin 3 (and 4) currents,
and the derivations of the anomalies. 
The first equation is the spin 3 generalization of the 
covariant form of gauge
(or gravitational) anomaly while the second is that of the
trace anomaly. 
The r.h.s. of the first equation consists of classical and quantum 
parts. The classical parts show that the electric and the 
gravitational backgrounds are not invariant under gauge transformations
associated with the spin 3 currents. 
The quantum part is the spin 3 generalization of 
the gauge anomalies  
in the electric and gravitational backgrounds.

For spin 4 current, they are given by
\begin{align}
 \nabla^\mu J^{(4)}_{\mu\nu\rho\sigma} =&  
  F_{\mu\nu} {J^{(3)\mu}}_{\rho \sigma}
  + F_{\mu\rho} {J^{(3)\mu}}_{ \sigma\nu}
  + F_{\mu\sigma} {J^{(3)\mu}}_{\nu\rho } 
  -\frac{1}{8}R
  \left(  
   \nabla_\nu J^{(2)}_{\rho\sigma} 
   + \nabla_\rho J^{(2)}_{\sigma\nu}
   + \nabla_\sigma J^{(2)}_{\nu\rho} 
  \right) 
  \nonumber \\
 & -\frac{1}{6}
  \left( 
   J^{(2)}_{\nu\rho} \nabla_\sigma R 
   +J^{(2)}_{\rho\sigma} \nabla_\nu R
   + J^{(2)}_{\sigma\nu} \nabla_\rho R 
  \right) 
  \nonumber \\
 & -\frac{1}{24} 
  \Bigl( 
  J^{(1)}_{\nu}\nabla_\rho \nabla_\mu {F^\mu}_{\sigma} 
  +J^{(1)}_{\rho}\nabla_\sigma \nabla_\mu {F^\mu}_{\nu}
  +J^{(1)}_{\sigma}\nabla_\nu \nabla_\mu {F^\mu}_{\rho} 
  \nonumber \\
 & +J^{(1)}_{\rho}\nabla_\nu \nabla_\mu {F^\mu}_{\sigma}
  +J^{(1)}_{\nu}\nabla_\sigma \nabla_\mu {F^\mu}_{\rho}
  +J^{(1)}_{\sigma}\nabla_\rho \nabla_\mu {F^\mu}_{\nu}  
  \Bigr) 
  \nonumber \\
& -\frac{\hbar}{960\pi}\frac{c_R-c_L}{2}
\Bigl(
\epsilon_{\nu\alpha} \nabla^\alpha \nabla_\rho \nabla_\sigma R+
\epsilon_{\rho\alpha} \nabla^\alpha \nabla_\sigma \nabla_\nu R+
\epsilon_{\sigma\alpha} \nabla^\alpha \nabla_\nu \nabla_\rho R
 \Bigr)  \nonumber \\
 & -\frac{1}{4}
  \left(
   g_{\nu\rho}\hat{C}_{\sigma}+g_{\rho\sigma}\hat{C}_{\nu}
   +g_{\sigma\nu}\hat{C}_{\rho}  \right),
\label{conservation spin 4}
\end{align} 
 \begin{align} 
 {J^{(4)\mu }}_{\mu \nu\rho}
  =-\frac{\hbar}{160\pi}\frac{c_L+c_R}{2} \nabla_\nu \nabla_\rho R
  +g_{\nu\rho}\frac{c_L+c_R}{2}\left[
  \frac{\hbar}{160\pi}\nabla^2 R+\frac{\hbar}{24\pi}\left({\tilde{F}}^2
  -\frac{13}{120}R^2  \right)  \right] 
 \end{align} 
where  $\hat{C}_\nu$ is defined as
\begin{align} 
 \hat{C}_{\nu} \equiv&  -\frac{1}{4}R \nabla_\rho {J^{(2)\rho}}_{\nu} 
 -\frac{1}{3} {J^{(2)\rho}}_{\nu} \nabla_\rho R 
 -\frac{1}{12} \Biggl(  J^{(1)\rho}\nabla_\rho \nabla_\mu {F^\mu}_{\nu}
 +J^{(1)}_{\rho}\nabla_\nu \nabla_\mu F^{\mu\rho}  \Biggr) \nonumber \\
 &-\frac{\hbar}{960\pi}\frac{c_R-c_L}{2}
\left(   \epsilon_{\nu\alpha} \nabla^\alpha \nabla_\rho \nabla^\rho R+
2\epsilon_{\rho\alpha} \nabla^\alpha \nabla^\rho \nabla_\nu R     \right) .
\end{align} 
Here we notice that these forms of anomalies are
specific to fermions. 
In the case of bosons, or fields with different conformal
weights, they have different forms.

\section{Anomaly method for  Hawking radiation}
\setcounter{equation}{0} 
\label{5.0} 
We now apply our results summarized in the previous section to
Hawking radiation in black hole backgrounds. 
In black hole backgrounds, by
using an appropriate harmonic expansion, one can describe matter fields near
the horizon by an infinite set of two-dimensional fields in the $(t, r)$
section of spacetime. Further since mass and potential terms
are suppressed in the near-horizon region and kinetic terms dominate the
action there, the theory becomes conformal near the horizon \cite{conformal}
and we can employ conformal field theory techniques.
Hence the following analysis of Hawking radiation
can be applied to any dimensional black holes, not
restricted to d=2.

In \cite{IMU2}, we derived fluxes of the HS currents in a
Schwarzschild black hole and showed that they coincide with the 
higher moments of
the thermal distribution. We there used transformation properties of the
holomorphic currents under a conformal transformation which relate
quantities near the horizon with those at infinity, and imposed the
regularity condition at the horizon. Further this analysis was extended to
charged black holes by considering gauge transformations in addition to
conformal transformations \cite{IMU3}. 
The above method can successfully reproduce the 
thermal radiation with the correct Hawking temperature, but
it will be instructive to obtain the same results
by  direct applications of the anomaly equations in the previous
section. In this section, we give a direct generalization of 
the anomaly method \cite{Robinson, IUW1} to spin 3 and 4 
fluxes.

In the  paper \cite{Robinson}, only the consistent
form of the energy-momentum tensor is used, but
in \cite{IUW1}, we showed that the boundary condition
at the horizon should be imposed  for the covariant current.
See the appendix of  \cite{IUW2} for why 
the boundary condition employed in \cite{Robinson}
does not work for the charged black holes. 
In the following discussions, 
we adopt the method using only the covariant currents \cite{Banerjee} 
because the anomaly equations in the previous section are
written  in the covariant form and we do not discuss
the consistent currents. 

\subsection{Spin 3 gauge current}
First let us consider the spin 3 current.
We divide the region outside the horizon into 
near-horizon region $H$ ($r \in [r_H, r_+ + \epsilon]$)  and 
the other region $O$  ($r > r_+ + \epsilon$).  
If we omit the ingoing modes near the horizon, the currents have the 
anomalies there. Thus the covariant current ${J^{(3)\mu}_{(H)}}_{\nu\rho}$ 
in the near-horizon region satisfies eq. (\ref{conservation spin 3}) with $c_L=0, c_R=1$.  
In the following, we put $\hbar=1$ for 
notational simplicity. $\hbar$ can be recovered by replacing
$(1/\pi)$ by $(\hbar/\pi)$ in the anomaly equations.

Since the expectation value of the current depends only on $r$ in the
black hole background considered here, 
the $\nu=\rho=t$ component of the equation 
(\ref{conservation spin 3}) becomes
\begin{eqnarray}
 \langle \partial_r {J^{(3)r}_{(H)}}_{tt} \rangle 
  &=& -2F_{tr}\langle {J^{(2)r}_{(H)}}_t \rangle
  - \frac{1}{8} \nabla_t \left(R \langle {J^{(1)}_{(H)}}_t \rangle \right) 
  + \frac{1}{16} g_{tt} \nabla_\mu 
  \left(R \langle{J^{(1)\mu}_{(H)}}\rangle \right)
  \nonumber \\
  && + \frac{1}{98\pi}
   \left(
    -2\nabla^r \nabla_\mu {F^\mu}_t 
    - g_{tt} \epsilon_{\alpha\sigma}\nabla^\sigma\nabla_\mu F^{\mu\alpha}
   \right).
   \label{spin3 region H}
\end{eqnarray} 
Here $\langle {J^{(1)r}_{(H)}}\rangle$ and
$\langle{J^{(2)r}_{(H)}}_t\rangle$ are expectation values of the covariant
$U(1)$ current and the covariant EM tensor in the near-horizon region,
respectively. They are determined by solving eqs. (\ref{conservation spin 1}) and (\ref{conservation spin 2}) with 
$c_L=0, c_R=1$ and requiring that they vanish at the horizon.
The results are 
\begin{eqnarray}
 \langle{J^{(1)r}_{(H)}}\rangle 
  &=& \frac{1}{2\pi}\left(A_t(r) - A_t(r_+)\right), \\
 \langle {J^{(2)r}_{(H)}}_t \rangle &=& \int^r_{r_+}dr \partial_r
  \left[
   -\frac{1}{2\pi} A_t(r_+)A_t(r) + \frac{1}{4\pi}A_t(r)^2
   + \frac{1}{96\pi} \left(ff''-\frac{1}{2}f'^2\right)
  \right].
\end{eqnarray}
See \cite{IUW1} and \cite{Banerjee} for the details.

Eq. (\ref{spin3 region H}) can be solved 
by using the above solutions for spin 1 and 2 currents
and imposing the vanishing condition 
at the horizon, $\langle {J^{(3)r}_{(H)}}_{tt} (r_+)\rangle =0$.
This vanishing condition is imposed for the covariant 
current, which is required by the regularity 
of the physical quantities at the horizon.
Now $\langle {J^{(3)r}_{(H)}}_{tt} \rangle$ is obtained as
\begin{eqnarray}
 \langle {J^{(3)r}_{(H)}}_{tt} \rangle &=& \int^r_{r_+} dr \partial_r
  \left[
   2c^{(2)}_O A_t + c^{(1)}_O A_t(r)^2 
   + \frac{1}{16}c^{(1)}_O (ff''-f'^2)
   + \frac{1}{6\pi}A_t(r)^3 
   \right. \nonumber \\
 && \hspace{12mm}
  \left.
   - \frac{1}{96\pi}f ^2 \partial_r^2 A_t(r)
   + \frac{1}{48\pi}ff'\partial_r A_t(r)
   + \frac{1}{32\pi}(ff''-f'^2)A_t(r)
  \right],
\end{eqnarray}
where $c^{(1)}_O$ and $c^{(2)}_O$ are given by
\begin{eqnarray}
 c^{(1)}_O &=& -\frac{1}{2\pi}A_t(r_+) = \frac{Q}{2\pi r_+}, \\
 c^{(2)}_O &=& \frac{1}{4\pi}A_t(r_+)^2 + \frac{1}{192\pi}f'(r_+)^2
  = \frac{Q^2}{4\pi r_+} + \frac{\kappa^2}{48\pi}.
\end{eqnarray}
They are  the 
the fluxes of the $U(1)$ current and
the EM tensor  at infinity.

In the outside region, the current ${J^{(3)r}_{(O)}}_{tt}$ has no
anomaly. Then it is given by
\begin{eqnarray}
 \langle {J^{(3)r}_{(O)}}_{tt} \rangle &=& c^{(3)}_O + 2c^{(2)}_O A_t 
  + c^{(1)}_O A_t(r)^2 
  + \frac{1}{16}c^{(1)}_O (ff''-f'^2)
  + \frac{1}{6\pi}A_t(r)^3, 
\end{eqnarray}
where the integration constant
$c^{(3)}_O$ gives the value of the flux at infinity. 
The current is
written as a sum in two regions
\begin{equation}
 {J^{(3)r}}_{tt}
  = {J^{(3)r}_{(O)}}_{tt}\Theta_+(r) + {J^{(3)r}_{(H)}}_{tt}H(r),
\end{equation}
where $\Theta_+(r) = \Theta(r - r_+ - \epsilon)$ and $H(r)=1-\Theta_+(r)$
are step functions which are defined in the region $r \geq r_+$.  Note that
this current is a part of the total current because we omitted the ingoing
modes near the horizon. The contribution from the ingoing modes $\langle
{K^{(3)r}}_{tt} \rangle$ is determined by imposing the following
two conditions: 
(1) it cancels the anomalous part in the near-horizon region,  
(2) there is no ingoing flux from the infinity.  
The second condition is required in the
Unruh vacuum. 
Then we have
\begin{eqnarray}
 \langle {K^{(3)r}}_{tt} \rangle &=& 
  -\left[
    \frac{1}{6\pi}A_t(r)^3
    - \frac{1}{96\pi}f ^2 \partial_r^2 A_t(r)
    + \frac{1}{48\pi}ff'\partial_r A_t(r)
    \right. \nonumber \\
 && \left. \hspace{8mm}
     + \frac{1}{32\pi}(ff''-f'^2)A_t(r)
    \right] H(r). 
\end{eqnarray}
Requiring the continuity  of the total current
at $r = r_+ + \epsilon$, 
which contains both of the outgoing and ingoing modes,   
the asymptotic flux of the  spin-3 current is obtained as 
\begin{eqnarray}
 c^{(3)}_O = - \frac{\kappa^2}{24\pi}A_t(r_+) 
  - \frac{1}{6\pi} A_t(r_+)^3
  = \frac{\kappa^2}{24\pi} \frac{Q}{r_+} 
  + \frac{1}{6\pi} \left(\frac{Q}{r_+}\right)^3.
\end{eqnarray}
This value coincides with the $n=2$ moment of the 
Hawking fluxes in (\ref{FDdistribution}).

\subsection{Spin 4 gauge current}

By using a similar procedure, we can derive the flux of the spin 4 current.
The current in the near-horizon region ${J^{(4)r}_{(H)}}_{ttt}$ satisfies the
anomaly equation (\ref{conservation spin 4}) with $c_L=0, c_R =1$, 
and in the outside region,  
the current ${J^{(4)r}_{(O)}}_{ttt}$ 
satisfies the conservation equation without
the anomalies. 
By solving these equations with the regularity condition 
at the horizon, we have the following results
for the expectation values of the currents 
$\langle {J^{(4)r}_{(H)}}_{ttt} \rangle$ in region $H$,
$\langle {J^{(4)r}_{(O)}}_{ttt} \rangle$ in region $O$, 
and the contribution from the ingoing
modes $\langle {K^{(4)r}}_{ttt} \rangle$ :
\begin{eqnarray}
 \langle {J^{(4)r}_{(H)}}_{ttt} \rangle
  &=& \int^r_{r_+} dr \partial_r
  \left[
   3c^{(3)}_O A_t(r) + 3c^{(2)}_O A_t(r)^2 + c^{(1)}_O A_t(r)^3
   + \frac{1}{16} c^{(2)}_O \left(4ff''-5f'^2\right)
   \right. \nonumber \\
 && - \frac{1}{16\pi}c^{(1)}_O 
  \left(f^2\partial_r^2 A_t(r) -2ff'\partial_r A_t(r)
   +\left(-4ff''+5f^2\right)A_t(r) \right) 
  + \frac{1}{8\pi}A_t(r)^4
  \nonumber \\
 && 
  -\frac{1}{32\pi}f^2\partial_r^2A_t(r) A_t(r)
  + \frac{1}{16\pi}ff'\partial_r A_t(r) A_t(r)
  + \frac{1}{64\pi} \left(4ff''-5f'^2\right) A_t(r)^2
  \nonumber \\
 && \left.
     + \frac{1}{30720\pi}
     \left(
      43f'^4 - 112ff'^2f'' + 52f^2f''^2 + 36 f^2f'f''' -12 f^3f''''
     \right)
  \right], \\
 \langle {J^{(4)r}_{(O)}}_{ttt} \rangle
  &=& c^{(4)}_O + 3c^{(3)}_O A_t(r) 
  + 3c^{(2)}_O A_t(r)^2 + c^{(1)}_O A_t(r)^3
   + \frac{1}{16} c^{(2)}_O \left(4ff''-5f'^2\right)
   \nonumber \\
 && - \frac{1}{16\pi}c^{(1)}_O
  \left(f^2\partial_r^2 A_t(r) -2ff'\partial_r A_t(r)
   +\left( -4ff''+5f^2 \right)A_t(r) \right), \\
 \langle {K^{(4)r}}_{ttt} \rangle &=& 
  -\left[
    \frac{1}{8\pi}A_t(r)^4
    -\frac{1}{32\pi}f^2\partial_r^2A_t(r) A_t(r)
  + \frac{1}{16\pi}ff'\partial_r A_t(r) A_t(r)
  \right. \nonumber \\
 && 
     + \frac{1}{64\pi} \left(4ff''-5f'^2\right) A_t(r)^2
     \nonumber \\
 && \left.
     + \frac{1}{30720\pi}
     \left(
      43f'^4 - 112ff'^2f'' + 52f^2f''^2 + 36 f^2f'f''' -12 f^3f''''
     \right)
   \right].
\end{eqnarray}
By imposing the continuity of the total current, we
have the flux $c^{(4)}_O$ at infinity
\begin{equation}
 c^{(4)}_O =
 \frac{7\kappa^4}{1920\pi} 
 + \frac{\kappa^2}{16\pi}\left(\frac{Q}{r_+}\right)^2
 + \frac{1}{8\pi}\left(\frac{Q}{r_+}\right)^4.
\end{equation}
 The value of the flux $c^{(4)}_O$ coincides with the $n=3$
moment of the Hawking flux \ref{FDdistribution}.

We have shown that the values of the anomalies 
for spin 3 and 4 currents at the horizon determine the
Hawking fluxes at infinity. This derivation is a direct
generalization of the anomaly method \cite{Robinson,IUW1}
to HS fluxes.
Finally we notice that since we have used the anomaly
equations for HS currents constructed from fermions,
these results are valid only for fermions.
For other fields than fermions with a different
conformal weight, the relations between 
holomorphic and covariant HS currents in \cite{IMU4}
are changed and accordingly so are the anomaly equations.
It is an interesting problem to classify the 
anomaly equations and see the universality 
of Hawking radiation.

\section{Christen-Fulling method for HS currents}
\label{sec CF}
\setcounter{equation}{0} 
In this section, we show that we can also obtain 
the correct Hawking fluxes for the HS currents
by solving the conservation  equations and the 
trace anomalies
 for $c_L=c_R=1$.
It is a  generalization of Christensen and
Fulling's method \cite{CF}, in which the energy flux is obtained
by solving the conservation of EM tensor and trace anomaly.
In the case of charged black holes, we need to 
combine to solve the equations for the chiral $U(1)$ anomaly with
the trace anomaly \cite{IMU3}.
\subsection{Spins up to 4 }
Now let us solve the conservation equations and
 trace anomaly equations summarized 
in  section 2 for $c_L=c_R=1$.
In the following we  
employ the conformal gauge $ds^2=e^{\varphi} du dv$
for the gravitational background and the Lorenz gauge $\nabla^\mu A_\mu=0$
for the background gauge field.
In these gauges, the equations are decomposed into
holomorphic (i.e. independent of $v$)
and anti-holomorphic (indep. of $u$) components and
can be  solved as
 \begin{align}
  J^{(1)}_u=j^{(1)}(u)+\frac{1}{\pi} A_u,
  \label{spin-1-cov-hol}
 \end{align}
 \begin{align}
 J_{uu}^{(2)}  
  = j^{(2)}(u)
  +2A_u J_u^{(1)} -\frac{1}{\pi} A_u^2 
  + \frac{1}{24\pi}
  \left( 
  \partial_u^2\varphi-\frac{1}{2}\left( \partial_u \varphi \right)^2 
  \right),
 \end{align}
 \begin{align}
J_{uuu}^{(3)}  
 =& j^{(3)}(u)
 +4A_u J_{uu}^{(2)}
 +\frac{1}{4} \partial_u \varphi \partial_u J_u^{(1)} 
 +\left(
 -4A_u^2+\frac{1}{4}\left(\partial_u^2 \varphi
 -\left( \partial_u \varphi \right)^2 \right)  
 \right) J_u^{(1)} \nonumber \\
 & -\frac{1}{6\pi}
 \left( \partial_u^2 \varphi -\frac{1}{2}(\partial_u \varphi)^2\right) A_u 
 -\frac{1}{12\pi}\partial_u^2 A_u +\frac{4}{3\pi}A_u^3,
\end{align}
\begin{align}
 J^{(4)}_{uuuu}
  =& j^{(4)}(u)  
  +6A_u J^{(3)}_{uuu} +\frac{3}{4}\partial_u \varphi \partial_u J^{(2)}_{uu}
  + \left[
     \partial_u^2 \varphi -\frac{5}{2}  (\partial_u \varphi)^2
     -12 A_u^2
    \right] J^{(2)}_{uu}
  \nonumber \\
 & -\frac{3}{2} A_u \partial_u \varphi \partial_u J^{(1)}_{u}
  + \left[
    - \frac{3}{2} A_u \left(\partial_u^2 \varphi 
	    -  (\partial_u \varphi)^2\right)
     -\frac{1}{2} \partial_u^2 A_u 
     + 8 A_u^3
    \right] J^{(1)}_{u}
  \nonumber \\
 & +\frac{1}{2\pi}A_u 
\left[ \partial_u^2 A_u + \left(\partial_u^2 \varphi-\frac{1}{2}(\partial_u \varphi)^2 \right) A_u  -4 A_u^3 \right]  \nonumber \\ 
& -\frac{1}{160\pi}\left( \partial_u^4 \varphi - \partial_u \varphi \partial_u^3 \varphi + \frac{4}{3}(\partial_u^2 \varphi)^2 -
\frac{7}{3}( \partial_u \varphi)^2 \partial_u^2 \varphi
+ \frac{7}{12}( \partial_u \varphi)^4 \right) 
.
  \label{spin4-hol-cov}
\end{align}
Here $j^{(n)}(u)$ are arbitrary holomorphic functions,
which can be fixed by boundary conditions at the horizon.

 First, we consider the flux of the $U(1)$ current
$\langle J^{(1)}_u \rangle$ in the black hole background. 
We impose the regularity condition that 
$ \langle J^{(1)}_U \rangle$ is regular at the horizon $U=0$ in the Kruskal 
coordinate (\ref{UVtransf}).  
This condition
corresponds to taking the Unruh vacuum.
The current $J^{(1)}_u$ in the $(u,v)$ coordinates
and another $J^{(1)}_U$ in the Kruskal $(U,V)$ coordinates
are related by
\begin{align}
 J^{(1)}_{U} =-\frac{1}{\kappa U} J^{(1)}_u. 
\end{align}
Therefore the regularity condition for $\langle J^{(1)}_U \rangle$ means 
$\langle J^{(1)}_u \rangle \rightarrow 0$ at the horizon.  
Since $J^{(1)}_u$ is written as 
\begin{align}
 J^{(1)}_{u}=j^{(1)}(u)+\frac{e^2}{\pi}A_u,
\label{U(1)curr2}
\end{align}
the condition is equivalent to the following boundary condition for the
holomorphic current $j^{(1)}(u)$,
\begin{align} 
 \langle j^{(1)}(u) \rangle \mathop{\longrightarrow}^{r\rightarrow r_+} 
 - \frac{e^2}{\pi} A_u\Big|_{r=r_+}
 =\frac{e^2 Q}{2\pi r_+}.
\end{align} 
Besides, the Reissner-Nordstr\"om black hole solution is static and thus
$\langle j^{(1)}(u) \rangle$, which is a holomorphic function of $u=t-r_*$,
should be a constant, namely $\langle j^{(1)}(u) \rangle = e^2 Q/2\pi r_+$.
Therefore the flux at infinity is obtained as
\begin{align} 
\langle J^{(1)}_u \rangle \ \mathop{\longrightarrow}^{r\rightarrow \infty} \ 
 \langle j^{(1)}(u)\rangle =\frac{e^2 Q}{2\pi r_+}.
\end{align} 
This value is equal to the $n=0$ result in (\ref{FDdistribution}).

Similarly, we can derive fluxes of the higher-spin currents by imposing
the regularity condition that the currents in the Kruskal coordinate are
regular at the horizon. Also it is important that the expectation value of
the covariant HS current $\langle J^{(n)}_{u \cdots u} \rangle$ coincides
with the value of the conformal HS current 
$\langle j^{(n)} (u)\rangle$ 
at infinity because contributions from the background fields
approach to zero there.  The results for the HS currents up to
spin 4 are given as follows:
\begin{eqnarray}
 \langle J^{(2)}_{uu} \rangle 
  \ \mathop{\longrightarrow}^{r\rightarrow\infty} \
  \langle j^{(2)}(u) \rangle &=& 
  \frac{\kappa^2}{48\pi} + \frac{1}{4\pi}\left(\frac{eQ}{r_+}\right)^2, \\
 \label{spin 3 flux}
 \langle J^{(3)}_{uuu} \rangle 
  \ \mathop{\longrightarrow}^{r\rightarrow\infty} \
  \langle j^{(3)}(u) \rangle &=& 
  \frac{\kappa^2}{24\pi}\frac{eQ}{r_+} 
  + \frac{1}{6\pi}\left(\frac{eQ}{r_+}\right)^3, \\
 \label{spin 4 flux}
   \langle J^{(4)}_{uuuu} \rangle 
  \ \mathop{\longrightarrow}^{r\rightarrow\infty} \
  \langle j^{(4)}(u) \rangle &=& 
  \frac{7 \kappa^4}{1920\pi} 
  + \frac{\kappa^2}{16\pi}\left(\frac{eQ}{r_+}\right)^2
  + \frac{1}{8\pi}\left(\frac{eQ}{r_+}\right)^4 .
\end{eqnarray}
 These values  coincide with those in eq.
(\ref{FDdistribution}). 
We have defined the currents to be totally symmetric,
and their trace anomalies vanish 
at $r \rightarrow \infty $. Hence 
the fluxes of currents at infinity
are given by 
$J^{(n) r}_{t \cdots t}=J^{(n)}_{u \cdots u} - J^{(n)}_{v \cdots v}$.
In the Unruh vacuum, the ingoing modes vanish there and
we have
 $J^{(n) r}_{t \cdots t}=J^{(n)}_{u \cdots u}$. 
 
We have now shown that, by solving the conservation equations
and the trace anomalies up to spins 4, 
we can reproduce the correct results of higher-moments of
the Hawking fluxes.

\subsection{General higher spins}
For spins up to 4, we solved the conservation equations 
and trace anomalies in section 2 and obtained
equations (\ref{spin-1-cov-hol})-(\ref{spin4-hol-cov}).
There  
 holomorphic functions  $j^{(n)}(u)$  
  have been introduced in solving the
 equations in the form of  $\partial_v (\cdots)=0$.
They can be  determined by imposing
appropriate boundary conditions for the currents.  
In conformal field theories, 
these holomorphic functions $j^{(n)}(u)$ 
play an important role
as generators of conformal $W_{1+\infty}$ transformations which are
HS generalizations of conformal transformations. 
We call them holomorphic HS currents.
In \cite{IMU4}, we have derived 
a relation between these holomorphic HS currents
and the original covariant HS currents 
as an identity of their generating functions. 
Actually, the gauge and trace anomalies for spin 3 and 4
in section 2 were derived from these relations.
In this sense, solving the conservation equations 
and trace anomalies is equivalent to the relation
between the holomorphic and covariant HS currents.
In this subsection, we use
the relation to derive all the higher moments of 
Hawking fluxes. 

We first summarize the relation. See \cite{IMU4} for details.
A generating function of the 
holomorphic HS currents is defined as 
\begin{eqnarray}
 G_{hol}(u+\alpha, u-\alpha) 
  &\equiv& \sum_{n=0}^{\infty} \frac{(2i\alpha)^n}{n!} 
   j^{(n+1)}(u) = \Psi^\dagger(u+\alpha)\Psi(u-\alpha) + 
   \frac{i}{4 \pi \alpha} .
   \label{holomorphic generating function}
\end{eqnarray}
Here $\Psi(u)$
is a holomorphic fermion field, which is a
holomorphic function of $u$ in the electric 
and gravitational background and related to the
original fermion field $\psi(u,v)$.
Each HS current should be  regularized appropriately
and (\ref{holomorphic generating function})
should be considered as a formal power series of $\alpha$.

By using the definition of the holomorphic fermion field 
$\Psi(u)$, this generating function can be written 
as 
\begin{eqnarray}
 G_{hol}(u+\alpha, u-\alpha) 
  &=& e^{\frac{1}{2}(\varphi(u+\alpha)+\varphi(u-\alpha))
  -2i \left(\eta(u+\alpha) - \eta(u-\alpha)\right)} 
  \nonumber \\
 && \times\Biggl[
  e^{-\frac{1}{4}(\varphi(u+\alpha)+\varphi(u-\alpha))
  +i \left(\eta(u+\alpha) - \eta(u-\alpha)\right)}
  \psi^\dagger(u+\alpha, v)\psi(u-\alpha, v) 
  \nonumber \\
 && \hspace{5mm} 
  +\frac{i\hbar}{2\pi}\frac{1}{x(u+\alpha)-x(u-\alpha)}   \Biggr]
  \nonumber \\
 && -\frac{i\hbar}{2\pi}
  \frac{ e^{\frac{1}{2}(\varphi(u+\alpha)+\varphi(u-\alpha))
  -2i \left(\eta(u+\alpha) - \eta(u-\alpha)\right)  }}
  {x(u+\alpha)-x(u-\alpha)}
  +\frac{i\hbar}{4\pi \alpha}.
\label{Ghol-cov}
\end{eqnarray}
Here we formally introduced a new coordinate $x$ 
which satisfy  $\partial_x=e^{-\varphi(u,v)} \partial_u$
and regard $x$ as a function of $u$ (i.e., $v$ is kept 
fixed).  
$\varphi(u,v)$ is the conformal factor of the gravitational
background and $\eta(u,v)$ represents the electric background
in the Lorentz gauge, $A_u=\partial_u \eta(u,v).$
The first term in (\ref{Ghol-cov}) with a bracket
$[ \cdots ]$ 
 can be written in terms of the regularized
covariant HS currents and their derivatives. 
The second and third terms give quantum corrections.
By expanding the r.h.s. of (\ref{Ghol-cov})
with respect to the parameter $\alpha$, 
we can obtain the equations relating the
holomorphic HS currents with the sum of the covariant HS currents
and quantum corrections:
\begin{align}
 \sum_{n=0}^{\infty} \frac{(2i\alpha)^n}{n!} \left(
   J^{(n+1)}_{u\cdots u}+\sum_{m,k=0}^{n-1}  C_{mk}^{(n+1)}(\varphi,A_u) \partial_u^k J^{(m+1)}_{u\cdots u} 
   + \hbar D^{(n+1)}(\varphi, A_u)
   \right)
\end{align}
where $C_{mk}(\varphi,A_u)$ and $D^{(n+1)}(\varphi,A_u)$
are  functions of $\varphi$ and $A_u$ which
 vanish at infinity.
 By comparing the coefficients of $\alpha^n$ up to $n=3$,
we get  the equations in 
 (\ref{spin-1-cov-hol})-(\ref{spin4-hol-cov}).
 Hence the above relation contains the full information
 of the anomalies.
We use  it
to evaluate the fluxes of the HS currents than spin 4.

Let us calculate the Hawking fluxes.
First we evaluate the expectation value of (\ref{Ghol-cov})
 in the near horizon region.
We impose the regularity condition for the covariant currents
at the horizon; $J^{(n)}_{u \cdots u} \rightarrow 0$.
Then the first term in the r.h.s. of
 (\ref{Ghol-cov}) becomes $0$. 
Thus we need to evaluate the last two terms, which are
proportional to $\hbar$.

In order to evaluate these quantum contributions, we first note that
 $\partial_u=-f \partial_r /2$ and $f\rightarrow 0$ near the horizon
(see the appendix \ref{App-Hawking}).
Then we have, near the horizon $u \rightarrow \infty$, 
$\eta(u+\alpha)-\eta(u-\alpha) \rightarrow
 2 \alpha \partial_u \eta = 2 \alpha A_u(r_+)=- \alpha Q/r_+.$
Higher derivatives $\partial_u ^n \eta$ for $n>1$ vanish at the horizon.
Because of the same reason, we have
$\varphi(u+\alpha)+\varphi(u-\alpha) \rightarrow 2\varphi(u)$ and 
\begin{align}
& x(u+\alpha)-x(u-\alpha) = 2 \sum_{n=0}^\infty
 \frac{\alpha^{2n+1}}{(2n+1)!} \partial_u^{2n+1}x 
\nonumber
\\ & \rightarrow 2 \sum_{n=0}^\infty \frac{\alpha^{2n+1}}{(2n+1)!}
 (\partial_u \varphi)^{2n} e^\varphi 
= 2 \frac{e^\varphi}{\kappa} \sinh(\alpha \kappa)
\end{align}
where $\kappa$ is the surface gravity at the horizon;
 $\partial_u \varphi(u)= \partial_u \log f \rightarrow -\kappa$.
As a result, we obtain
\begin{eqnarray}
\left\langle G_{hol}(u+\alpha, u-\alpha) \right\rangle
 = \sum_{n=0}^{\infty} \frac{(2i\alpha)^n}{n!} 
  \left\langle  j^{(n+1)}(u)\right\rangle 
 \rightarrow  -\frac{i \hbar}{4\pi \alpha}
  \frac{\alpha \kappa e^{-\frac{2i \alpha Q}{ r_+}}}{
\sinh(\alpha \kappa)    }+\frac{i \hbar}{4\pi \alpha}.
\label{FDtwo-point}
\end{eqnarray}
By expanding this equation with respect $\alpha$ and comparing their 
coefficients, we can obtain the expectation value 
$\left\langle j^{(n)}(u) \right\rangle$ which 
are equal to the value of the covariant current 
$\left\langle J^{(n)}_{u\cdots u} \right\rangle$ at infinity. 
We can show that these values of the fluxes coincide with 
(\ref{FDdistribution}).
Besides, as is pointed out in \cite{IMU2, IMU3}, this generating
function is the regular part of the fermion Green function, and  
(\ref{FDtwo-point}) is interpreted as the temperature-dependent part of
the thermal Green function with chemical potential. 
It means that the thermal excitation of the fields are introduced
through the quantum effect. 

By this analysis, 
we can also see, which was implicitly used in \cite{IMU2,IMU3}, 
that the fluxes of the covariant HS currents corresponding 
to the  physical quantities are
equal to those of the holomorphic currents at infinity, since both values of 
$\varphi$ and $A_u$  asymptotically decay to zero.

\section{Summary \label{conclusion}}
\setcounter{equation}{0}

In this paper, 
by using the higher-spin (HS) generalizations of
the gauge and trace anomaly equations,
we reproduced the HS fluxes of Hawking radiation.
 
 We first used the anomaly equations for spin 3 and 4 currents
to derive the Hawking fluxes and reproduced the correct 
values. This is a direct HS generalization of the gauge and
 gravitational anomaly method by Robinson and
Wilczek \cite{Robinson} and by some of us \cite{IUW1}.

We then solved the conservation equations and 
trace anomalies for the HS currents.
 This again reproduces the correct values of Hawking
fluxes for spin 3 and 4. 
This is now considered as
a HS generalization of the method by Christensen and 
Fulling \cite{CF}.
We further employed the generating function of the general HS currents 
and derived  the complete Fermi-Dirac distribution.

Finally we should notice the following.
The anomaly equations we have used are constructed
from the fermion currents, and 
they are modified for other fields with different
conformal weights. This gives a different value
of HS fluxes of Hawking radiation for bosons.


\appendix

\section{Reissner-Nordstr\"om black hole}
\setcounter{equation}{0}
\label{App-Hawking}

We summarize the basics of Reissner-Nordstr\"om black holes.  The metric and
the gauge potential of Reissner-Nordstr\"om black holes with mass $M$ and
charge $Q$ are given by
\begin{eqnarray}
 ds^2 &=& f(r)dt^2-\frac{1}{f(r)}dr^2-r^2d\Omega_2^2,\\
 A_t &=& -\frac{Q}{r}, \label{bg-gauge}
\end{eqnarray}
where 
\begin{equation}
 f(r) = 1-\frac{2M}{r}+\frac{Q^2}{r^2} = \frac{(r-r_+)(r-r_-)}{r^2}
 \end{equation}
 and the radius of outer (inner) horizon $r_\pm$ is given by
 \begin{equation}
 r_\pm = M \pm \sqrt{M^2-Q^2}.
\end{equation}
It is useful to define the  tortoise coordinate 
by solving $  dr_* = dr/f $  as
\begin{equation}
  r_* = r + \frac{1}{2\kappa_+}\ln \frac{|r-r_+|}{r_+}
  + \frac{1}{2\kappa_-}\ln \frac{|r-r_-|}{r_-}.
  \end{equation}
  Here the surface gravity at $r_\pm$ is given by
  \begin{equation}
  \kappa_\pm = \frac{1}{2}f'(r_\pm) = \frac{r_\pm - r_\mp}{2r_\pm^2}.
\end{equation}
In the following we consider a region near the outer horizon.
First we define the light-cone coordinates, 
$u=t-r_*$ and $v=t+r_*$. $u(v)$ are the 
outgoing (ingoing) coordinates and the metric 
in these coordinates becomes as
\begin{equation}
 ds^2 = f(dt^2 - dr_*^2) - r^2d\Omega^2 
 = fdudv - r^2d\Omega^2. 
 \label{uv}
\end{equation}
In order to investigate the physics near the outer horizon,
since $(u,v)$ coordinate is still singular at the horizon, 
it is important to introduce a regular coordinate,  
Kruskal coordinate, defined by the transformations
\begin{equation}
   U=-e^{-\kappa_+ u}, \qquad V=e^{\kappa_+ v}.
   \label{UVtransf}
\end{equation}
The metric becomes
\begin{equation} 
 ds^2  = \frac{r_+r_-}{\kappa_+^2}\frac{1}{r^2}e^{-2\kappa_+ r}
       \left(\frac{r_-}{r-r_-}\right)^{\frac{\kappa_+}{\kappa_-}-1}
       dUdV  - r^2d\Omega^2.
       \label{UV}
\end{equation}
If we restrict to see the two-dimensional $(r,t)$ section,
both of these coordinates (\ref{uv}), (\ref{UV}) have the forms of 
the conformal gauge
\begin{eqnarray}
 ds^2 = e^{\varphi(u,v)} dudv = e^{\varphi'(U,V)}dUdV.
\end{eqnarray}
The transformation (\ref{UVtransf}) is a conformal 
transformation from an asymptotically flat coordinate
to a regular coordinate near the horizon.

In the bulk of the paper we omit the subscript $+$ of the surface gravity
$\kappa_+$ at the outer horizon because $\kappa_-$ does not appear in our
analysis.

\section{Higher moments of Hawking fluxes}
\setcounter{equation}{0}
\label{App-Flux}
In this appendix, we
calculate the higher-spin fluxes of the Fermi-Dirac
distribution with Hawking temperature and the chemical potential.
We first note that  two-dimensional effective theories
near the horizon in various black holes 
have the same form as that in charged black holes. It is thus
 enough to consider Hawking fluxes in the Reissner-Nordstr\"om black hole.
Some basic facts of the Reissner-Nordstr\"om black hole are summarized in
the appendix \ref{App-Hawking}. Thermal radiation of  fermions  with charge
$q$ from Reissner-Nordstr\"om black hole  satisfies the Fermi-Dirac
distribution $N_q(\omega)$ with a chemical potential corresponding to the
value of the electric potential at the horizon,
\begin{equation}
 N_q(\omega) = \frac{1}{e^{\beta (\omega - \frac{qQ}{r_+})} + 1}. 
\end{equation}
Here $\beta = \frac{2\pi}{\kappa}$ is the inverse temperature.  From
this distribution function, we can evaluate the thermal fluxes of 
higher-spin currents at  infinity as
\begin{eqnarray}
 && \int_0^\infty \frac{d\omega}{2\pi}
  \left[\omega^n N_e(w) - (-\omega)^n N_{-e}(w)\right] 
  \nonumber  \\
 && \qquad
  = \sum_{m=1}^{\lceil \frac{n}{2} \rceil} 
  \left(\frac{eQ}{r_+}\right)^{n+1-2m}
  \frac{n!\left(1-2^{1-2m}\right) B_m \kappa^{2m}}{2\pi (2m)!(n-2m+1)!}
  + \frac{1}{2\pi (n+1)} \left(\frac{eQ}{r_+}\right)^{n+1}.
  \label{FDdistribution}
\end{eqnarray}
Here $B_m$ is the Bernoulli number $(B_1 = 1/6, ~B_2 = 1/30)$ and 
$\lceil x \rceil$ is the ceiling function, which returns the smallest
integer not less than $x$.  This integral is the $n$-th moment of the
thermal flux  containing  contributions from the fermion with charge
$e$ and its antiparticle with charge $-e$.


\end{document}